\documentclass[12pt, letterpaper]{article}
\usepackage[margin=1in,footskip=0.25in]{geometry}
\usepackage{amsmath}
\usepackage{amsthm}
\usepackage{amssymb}
\usepackage{graphicx}
\usepackage{extarrows}
\usepackage{rotating}
\usepackage{epstopdf}
\usepackage{float}
\usepackage{caption}
\usepackage{subcaption}
\usepackage{natbib}
\usepackage{orcidlink}
\usepackage[normalem]{ulem}
\usepackage{float, latexsym, times}
\usepackage{amsfonts, amstext}
\usepackage{multirow}
\usepackage{booktabs}
\usepackage{comment}
\usepackage{color}
\usepackage{xcolor}
\usepackage{hyperref}

\newtheorem{definition}{Definition}
\newtheorem{theorem}{Theorem}
\newtheorem{lemma}{Lemma}

\begin{document}
%\title{NCST: An Alternative Construction of the Multivariate Skew $t$ Distribution with Applications in Diagnostic Modeling} \vspace{.1mm}

% 
%\title{Flexible Modeling of Multivariate Skewed and Heavy-Tailed Data via a Non-Central Skew $t$ Distribution with Applications in Diagnostic Modeling} \vspace{.1mm}

\title{Flexible Modeling of Multivariate Skewed and Heavy-Tailed Data via a Non-Central Skew $t$ Distribution: Application to Tumor Shape Data} \vspace{.1mm}

\author{Abeer M. Hasan$^{1}$\footnote{Corresponding author. Email: amhasan1@ncat.edu} \orcidlink{0009-0000-6444-9980},
         Ying-Ju Chen$^{2}$ \orcidlink{0000-0002-6444-6859}\\$^1$Department of Mathematics and Statistics\\North Carolina A\&T State University, Greensboro, NC 27411, USA\\$^2$Department of Mathematics\\University of Dayton, Dayton, OH 45469, USA}

%\date{Received: 5/31/2025/ Accepted: date}
% The correct dates will be entered by the editor

\setcounter{equation}{0} \numberwithin{equation}{section}

\maketitle

\begin{abstract}
 We propose a flexible formulation of the multivariate non-central skew $t$ (NCST) distribution, defined by scaling skew-normal random vectors with independent chi-squared variables. This construction extends the classical multivariate $t$ family by allowing both asymmetry and non-centrality, which provides an alternative to existing skew $t$ models that often rely on restrictive assumptions for tractability. We derive key theoretical properties of the NCST distribution, which includes its moment structure, affine transformation behavior, and the distribution of quadratic forms. Due to the lack of a closed-form density, we implement a Monte Carlo likelihood approximation to enable maximum likelihood estimation and evaluate its performance through simulation studies. To demonstrate practical utility, we apply the NCST model to breast cancer diagnostic data, modeling multiple features of tumor shape. The NCST model achieves a superior fit based on information criteria and visual diagnostics, particularly in the presence of skewness and heavy tails compared to standard alternatives, including the multivariate normal, skew normal, and Azzalini's skew $t$ distribution. Our findings suggest that the NCST distribution offers a useful and interpretable choice for modeling complex multivariate data, which highlights promising directions for future development in likelihood inference, Bayesian computation, and applications involving asymmetry and non-Gaussian dependence. 

\vspace{1mm} 

\end{abstract}

\section{Introduction}\label{sec:intro}
Skewed and heavy-tailed data are common across numerous applied domains such as econometrics, environmental science, and risk analysis. For example, income distributions, housing prices, and insurance claims often exhibit both asymmetry and excess kurtosis, with tail behavior capturing meaningful extreme events, such as financial losses or contaminant spikes, that cannot be treated as outliers. The importance of flexible models has been widely recognized: in finance and economics, heavy-tailed models are crucial for assessing systemic risk and modeling extreme events in the equity market \citep{ibragimov2015heavy, guo2017heavy}; in insurance, claims data routinely display strong tail dependence \citep{ahmad2022new}; and in environmental science, extreme value patterns arise in flood peaks and pollutant concentrations \citep{merz2022understanding, gomez2019gumbel}. In the recent study of the skewed heavy-tailed vector autoregressive framework proposed by \citet{karlsson2023vector} emphasizes the need for multivariate models that accommodate both skewness and varying degrees of tail behavior across dimensions.

The skew-normal distribution, introduced by \citet{Azzalini1985}, has become a foundational model for asymmetric data. It augments the normal distribution with a skewness parameter, which allows for control of asymmetry while maintaining analytical tractability. However, the skew-normal distribution has limited ability to capture heavy tails and is therefore not suitable for modeling data that exhibits both skewness and leptokurtosis. 

To improve modeling flexibility, a number of generalizations of the skew normal distribution have been developed in addition to Azzalini's popular formula. One example of such expansion is the framework proposed by \cite{WANG2009533}, which defines skew-normal distributions by a generalized linear transformation of latent variables. This approach broadens the family's ability to model complex dependence structures and non-diagonal covariance behavior, extending the support beyond what is attainable through standard location-scale transformations. 

To overcome the skew-normal distribution's inability to model heavy tails, the skew $t$ distribution was developed as a more general alternative. The skew $t$ distribution combines the flexibility of heavy tails with the skewness parameter of the skew normal distribution family and has been widely developed in both theory and applications. Extension methods such as the extended skew $t$ (EST) \citep{ArellanoValle2010} and parameterizations studied in \citet{ARELLANOVALLE2013} have enhanced the adaptability of the $t$ distribution. Applications range from financial time series modeling \citep{Nakajima2012} to environmental monitoring \citep{TAGLE2020} and robust modeling of truncated or censored data \citep{GALARZA2022104944}. 

Despite these advances, most multivariate skew $t$ models assume centrality, that is, a zero mean structure or symmetry around the origin. This limits their expressiveness when data exhibit non-central behavior due to structural shifts or inherent asymmetry, which creates a practical gap, particularly in multivariate settings where asymmetry and location shifts interact. 

To address this, \citet{Hasan2013, HasanNingGupta2016} introduced a univariate non-central skew $t$ (NCST) distribution by scaling a skew-normal variable with an independent chi-squared random variable. Their construction captures both skewness and non-centrality, which makes it a promising model for real-world asymmetric and heavy-tailed data. However, a multivariate extension of this framework has not been fully developed. 

In this paper, we build on Hasan et al.'s univariate NCST formulation and propose a multivariate non-central skew $t$ distribution. Our model preserves the generative intuition of the classical multivariate $t$ distribution while introducing parameters that flexibly control both skewness and non-centrality. We derive key theoretical properties of the distribution, develop a Monte Carlo likelihood framework for inference, and demonstrate the model's empirical advantages through simulation and real data analysis.

\subsection{The Multivariate Skew Normal Distribution}
\cite{AzzaliniDallaValle1996} extended the univariate skew-normal distribution to the multivariate setting. A random vector $\mathbf{Z} \in \mathbb{R}^k$ is said to follow a multivariate skew-normal distribution, denoted by $\mathbf{Z} \sim \mathrm{SN}_k(\Omega, \alpha)$, if its density function is given by
\begin{equation}\label{eq:ASN}
        f_{\mathbf{Z}}(\mathbf{z}) = 2 \, \phi_k(\mathbf{z}; \boldsymbol{\Omega}) \, \Phi(\boldsymbol{\alpha}^\top \mathbf{z}), \quad \mathbf{z} \in \mathbb{R}^k,
\end{equation}
where \( \phi_k(\cdot; \boldsymbol{\Omega}) \) is the \( k \)-dimensional normal density with mean zero and covariance matrix \( \boldsymbol{\Omega} \), and \( \Phi(\cdot) \) is the standard normal c.d.f. The parameter $\alpha$ is referred to as a shape parameter, although the actual shape is regulated in a more complex way \citep{WANG2009533}.

A general location-scale version of this distribution is obtained by applying an affine transformation to $\mathbf{Z}$. Specifically, let 
\begin{equation}\label{eq:ASN2}
    \mathbf{X} = \boldsymbol{\xi} + \boldsymbol{\omega} \mathbf{Z},
\end{equation}
where \( \boldsymbol{\xi} \in \mathbb{R}^k \) is a location vector and \( \boldsymbol{\omega} = \mathrm{diag}(\omega_1, \ldots, \omega_k) \) is a diagonal scale matrix. The resulting density of \( \mathbf{X} \sim \mathrm{SN}_k(\boldsymbol{\xi}, \boldsymbol{\Omega}, \boldsymbol{\alpha}) \) is
\begin{equation}\label{eq:ASN3}
    f_{\mathbf{X}}(\mathbf{x}) = 2 \, \phi_k(\mathbf{x} - \boldsymbol{\xi}; \boldsymbol{\Omega}) \, \Phi \left( \boldsymbol{\alpha}^\top \boldsymbol{\omega}^{-1}(\mathbf{x} - \boldsymbol{\xi}) \right),
\end{equation}
where \( \boldsymbol{\Omega} = \boldsymbol{\omega} \boldsymbol{\Omega}_Z \boldsymbol{\omega} \), with \( \boldsymbol{\Omega}_Z \) being the correlation matrix associated with \( \mathbf{Z} \).

%To accommodate more flexible linear transformations beyond diagonal scaling, \cite{WANG2009533} proposed a generalized framework in which $\mathbf{Y} = \boldsymbol{\xi} + \mathbf{B}^\top\mathbf{Z}$, with $\mathbf{Z} \sim \mathrm{SN}_k(\boldsymbol{I_k, \alpha})$ and a general matrix $\mathbf{B} \in \mathbb{R}^{k \times n}$, is said to follow $\mathcal{SN}_n(\boldsymbol{\xi}, \mathbf{B}, \boldsymbol{\alpha})$. Azzalini's formulation is recovered as a special case when $\mathbf{B}=\Omega^{1/2}$. While Wang's extension enables broader covariance structures and transformation behavior, our NCST formulation builds on Azzalini's classical model as implemented in the \texttt{sn} package.

\subsection{Novelty and Contribution}
We propose a novel construction of the multivariate non-central skew $t$ (NCST) distribution by scaling a multivariate skew-normal random vector with an independent chi-squared variable. This generative approach preserves the intuitive structure of the classical multivariate $t$ distribution while explicitly incorporating both skewness and non-centrality. The resulting model is particularly useful when data exhibit asymmetric behavior not centered at the origin, an essential feature for many real-world applications. 

While prior skew-$t$ models, such as those in \citet{Gupta2003} and \citet{ARELLANOVALLE2013}, provided flexible representations of heavy-tailed, asymmetric data, most impose constraints such as zero-location parameters or symmetric dependence structures to ensure closed-form densities. These simplifications improve tractability but restrict modeling flexibility. In contrast, our NCST formulation allows for asymmetry, heavy tails, and non-zero centering without requiring such restrictions. 

From a methodological perspective, the lack of a closed-form likelihood necessitates a Monte Carlo approximation framework for likelihood-based inference. While such approaches are not new, we implement and assess this procedure in the context of the NCST model, illustrating its feasibility and identifying practical strategies to improve stability, such as initialization using simpler nested models. 

The utility of the NCST model is assessed through simulation and by applying it to a real-world dataset. By comparing with standard alternatives such as the multivariate normal, the skew-normal, and Azzalini's skew $t$ distribution, the advantage of our approach in capturing both tail behavior and skewness is highlighted, particularly under non-central configurations. 

In summary, this paper makes the following main contributions:
\begin{itemize}
    \item[1.] Proposes a natural and flexible construction of the multivariate NCST distribution that allows multiple tuning parameters to control the skewness and tail behavior;
    \item[2.] Establishes new theoretical results, including moment structure, affine transformations, and the distribution of quadratic forms;
    \item[3.] Implements likelihood-based inference using Monte Carlo approximation, with practical guidance for initialization and optimization;
    \item[4.] Demonstrates, through simulation and application, the empirical value of the NCST model in modeling asymmetric, heavy-tailed, non-central multivariate data. 
\end{itemize}

The remainder of this paper is organized as follows. Section \ref{sec:skew_t} introduces the multivariate NCST distribution and its construction. Section \ref{sec:theory} presents key theoretical properties. Section \ref{sec:simulation} presents results from simulation studies. Section \ref{sec:applications} applies the model to a real-world dataset. Section \ref{sec:conclusion} concludes with a discussion of the findings and future research directions.

%\subsection{Literature Review And Motivation}\label{subsec:literature}
%\input{02_Literature}

\section{A Multivariate Non-Central Skew \textit{t} Distribution}\label{sec:skew_t}
In this section, we present a multivariate generalization of Hasan's non-central skew $t$ distribution presented and discussed in the univariate case in \cite{Hasan2013, HasanNingGupta2016}.

\begin{definition}\label{de:NCST}\textbf{(Multivariate Non-Central Skew $t$ Distribution, (NCST))}\\
Let \( \boldsymbol{X} \sim \mathrm{SN}_k(\boldsymbol{\xi}, \boldsymbol{\Omega}, \boldsymbol{\alpha}) \) denote a $k$-dimensional multivariate skew-normal random vector in the sense of \citet{AzzaliniDallaValle1996}, with location vector \( \boldsymbol{\xi} \in \boldsymbol{R}^k \), positive definite covariance matrix \( \boldsymbol{\Omega} \in \boldsymbol{R}^{k \times k} \), and shape (skewness) vector \( \boldsymbol{\alpha} \in \boldsymbol{R}^k \). Let \( Y \sim \chi^2_r \) be independent of \( \boldsymbol{X} \).
Then the multivariate non-central skew-$t$ random vector \( \boldsymbol{T} \in \boldsymbol{R}^k \) is defined by:
\begin{equation} \label{eq:mvst1}
T_i = \frac{X_i}{\sqrt{Y/r}} , \quad i = 1, 2, \dots, k.
\end{equation}
We denote this distribution as $\boldsymbol{T} \sim \mathrm{NCST}_{k}(\boldsymbol{\xi}, \boldsymbol{\Omega}, \boldsymbol{\alpha}, r)$.
\end{definition}

The probability density function of the NCST distribution can be represented as a scale mixture of multivariate skew-normal densities. Specifically, letting $f_X(\cdot)$ denote the density of the $SN_k(\boldsymbol{\xi}, \boldsymbol{\Omega}, \boldsymbol{\alpha})$ and $f_Y(y)$ the density of the chi-squared distribution with $r$ degrees of freedom, the NCST density for a point $t\in \boldsymbol{R}^k$ can be expressed as:

\[
f_T(t) = \int_0^\infty f_X\left(t \cdot \sqrt{\frac{y}{r}}\right) \cdot \left(\sqrt{\frac{y}{r}}\right)^k\cdot f_Y(y) dy. 
\]

\begin{figure}[H]
    \centering
    \includegraphics[width=0.9\linewidth]{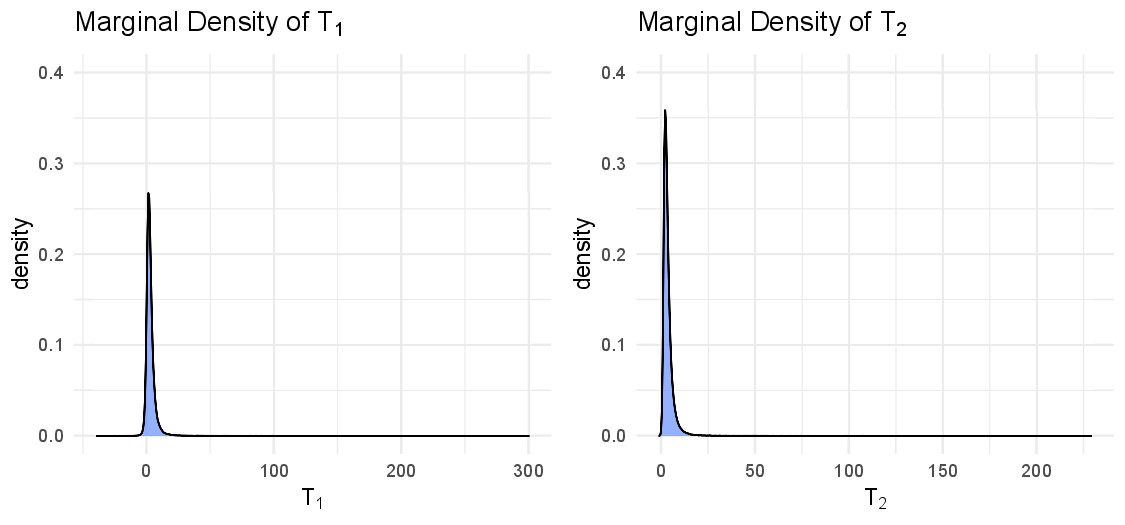}
    \caption{Marginal density estimates of $T_1$ and $T_2$ from the bivariate non-central skew $t$ distribution with parameters $\boldsymbol{\xi}=(1,2)^\top, \boldsymbol{\Omega} = diag(4,1), \boldsymbol{\alpha} = (3,3)^\top$, and degrees of freedom $r = 3$.}
    \label{fig:marginal_densities}
\end{figure}

\begin{figure}[H]
	\centering
	\begin{subfigure}{0.48\linewidth}
		\includegraphics[width=\linewidth]{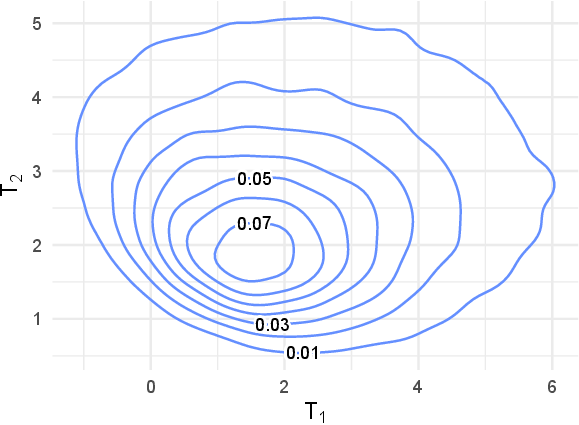}
	\end{subfigure}
	\begin{subfigure}{0.48\linewidth}
		\includegraphics[width=\linewidth]{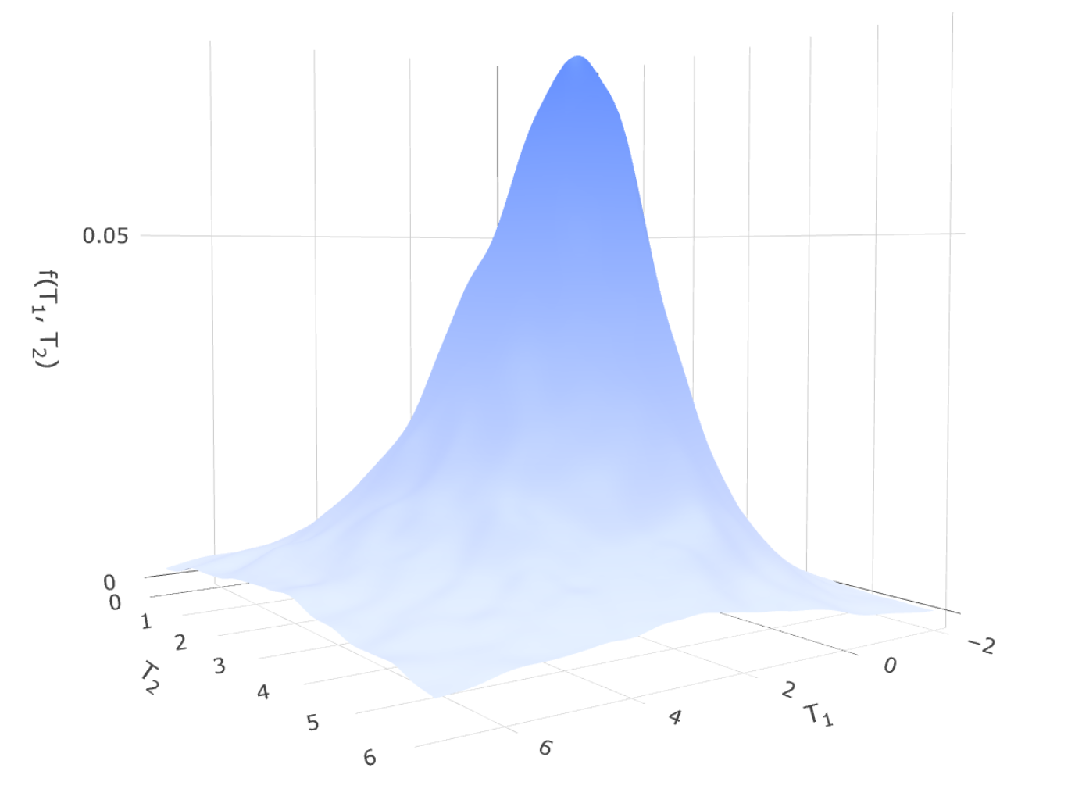}
	\end{subfigure}
	\caption{Contour (left) and 3D surface (right) plots of the estimated joint density $f(T_1, T_2)$ from a bivariate non-central skew $t$ distribution with parameters $\boldsymbol{\xi}=(1,2)^\top, \boldsymbol{\Omega} = diag(4,1), \boldsymbol{\alpha} = (3,3)^\top$, and degrees of freedom $r = 3$.}
	\label{fig:joint_density}
\end{figure}

Although this integral does not generally admit a closed-form expression, numerical integration or Monte Carlo approximation provides a tractable and accurate means for evaluating the density in practice. To illustrate the distributional characteristics of the proposed multivariate non-central skew $t$ distribution, we consider the bivariate case with parameters $\boldsymbol{\xi}=(1,2)^{\top}, \boldsymbol{\Omega} = diag(4,1), \boldsymbol{\alpha} = (3,3)^{\top}$, and degrees of freedom $r = 3$, where $^{\top}$ indicates the matrix transpose operator, as an example. Figure \ref{fig:marginal_densities} presents the marginal density estimates of $T_1$ and $T_2$, demonstrating the pronounced right skewness and heavy-tailed behavior induced by the non-zero skewness parameters and low degrees of freedom. Figure \ref{fig:joint_density} then displays the joint structure of $(T_1, T_2)$: the left panel shows the contour plot of the estimated joint density, providing a top-down view that highlights the asymmetry and directional concentration of the probability mass, while the right panel presents the corresponding 3D surface plot, which emphasizes the sharp peak and extended tails along the positive axes.

To contextualize the novelty and structure of our proposed NCST distribution, we provide a direct comparison with one of the most well-established alternatives: \cite{Azzalini_2013}'s multivariate skew $t$ distribution. While both formulations incorporate skewness and heavy tails through a scale-mixture framework involving the skew-normal distribution and a chi-squared variable, they differ substantially in how location, scale, and skewness are introduced and interact. Table \ref{tab:skewt_comparison} outlines these key differences across stochastic form, parameterization, tail behavior, and implementation. In contrast, Azzalini's model adds the location shift after scaling, which complicates the role of the location parameter. 

\begin{table}[H]
    \centering
        \caption{Comparison between the proposed non-central skew $t$ distribution and Azzalini's skew $t$ distribution \citep{Azzalini_2013}.}
    \label{tab:skewt_comparison}
    \begin{tabular}{p{4cm}|p{5.5cm}|p{5.5cm}}
    \toprule
    \textbf{Feature}    &  \textbf{Non-central Skew $t$} & \textbf{Azzalini's Skew $t$}\\
    \midrule
    Stochastic Form     &  $\boldsymbol{T} = \dfrac{\boldsymbol{X}}{\sqrt{Y/r}}$ \newline where $\boldsymbol{X} \sim SN_k(\boldsymbol{\xi}, \boldsymbol{\Omega}, \boldsymbol{\alpha})$, \newline $Y \sim \chi^2_r$ &
    $\boldsymbol{T} = \boldsymbol{\xi} + \dfrac{\boldsymbol{X}}{\sqrt{Y/r}}$ \newline where $\boldsymbol{X} \sim SN_k(\boldsymbol{0}, \boldsymbol{\Omega}, \boldsymbol{\alpha})$, \newline $Y \sim \chi^2_r$ \\
    \midrule
    Skewness source & Shape parameter $\boldsymbol{\alpha}$ inside \newline $\boldsymbol{X} \sim SN_k(\cdot)$ & Shape parameter $\boldsymbol{\alpha}$ inside \newline $\boldsymbol{X} \sim SN_k(\cdot)$ \\
    \midrule
    Non-centrality & Location shift $\boldsymbol{\xi}$ inside $\boldsymbol{X}$ before scaling & Location shift $\boldsymbol{\xi}$ added after scaling \\
    \midrule
    Tail behavior & Heavy tails via $Y \sim \chi^2_r$ & Heavy tails via $Y \sim \chi^2_r$ \\
    \midrule
    Covariance structure & Matrix $\boldsymbol{\Omega}$ inside $\boldsymbol{X}$ & Matrix $\boldsymbol{\Omega}$ inside $\boldsymbol{X}$ \\
    \midrule
    Implementation in R & Custom implementation & Available via \texttt{sn::mst()}, \texttt{sn::rst()}, \texttt{sn::selm()} \\
    \midrule
    Modeling flexibility & Allows separate control of location, skewness, and tail heaviness & Skewness and scale are in the core; location is added after scale\\
    \bottomrule
    \end{tabular}
\end{table}

\section{Theoretical Properties}\label{sec:theory}
This section presents key theoretical properties of the NCST distribution. We first derive general moments $\mathbb{E}[\boldsymbol{T}^k]$, which provide insight into how skewness and non-centrality affect the distribution's shape across different orders. These results go beyond standard location and scale summaries to characterize higher-order structures such as asymmetry and tail behavior. Next, we establish the behavior of the NCST distribution under linear transformations, showing that it retains its distributional form under certain mappings. Finally, we characterize the distribution of quadratic forms involving NCST random vectors, an important result for likelihood-based inference, projection statistics, and control chart applications. 

\subsection{Moments and Expectations}
This subsection derives analytical expressions for the marginal raw moments of the NCST distribution. Specifically, we compute $\boldsymbol{E}[T_i^k]$ for each component $T_i$ of the random vector $\boldsymbol{T}\sim \mathrm{NCST}_k(\cdot)$, by expressing them in terms of the corresponding moments of the underlying multivariate skew-normal and chi-squared distributions. These expressions allow us to characterize the tail behavior and skewness of each marginal distribution under varying degrees of freedom and asymmetry. Here, the notation $\boldsymbol{T}^k$ denotes the element-wise power, i.e., $\boldsymbol{T}^k=(T_1^k, \ldots, T_k^k)^\top$, and $\boldsymbol{E}[\boldsymbol{T}^k]$ refers to the vector of marginal $k^{\text{th}}$ moments. 

\begin{theorem} \label{th:kst}
Let $\boldsymbol{T} \sim \mathrm{NCST}_{k}(\boldsymbol{\xi}, \boldsymbol{\Omega}, \boldsymbol{\alpha}, r)$. Then, for any positive integer $k$, the $k^{th}$ raw moment of $\boldsymbol{T}$ is given by:
\begin{equation} \label{eq:kth}
\boldsymbol{E}[\boldsymbol{T}^k] = r^{k/2} \cdot \boldsymbol{E}[\boldsymbol{X}^k] \cdot \boldsymbol{E}[{Y}^{-k/2}],
\end{equation}
 where $\boldsymbol{X} \sim \mathrm{SN}_n(\boldsymbol{\xi}, \boldsymbol{\Omega}, \boldsymbol{\alpha})$, and $Y\sim \chi_r^2$, with $\boldsymbol{X}$ and $\boldsymbol{Y}$ independent.
\end{theorem}

\begin{proof}
The result follows immediately from the stochastic representation
\[
\boldsymbol{T} = \frac{\boldsymbol{X}}{\sqrt{Y/r}} = \sqrt{\frac{r}{Y}}\cdot \boldsymbol{X}
\]
and the independence of $\boldsymbol{X}$ and $Y$, which yields
\[
\boldsymbol{E}[\boldsymbol{T}^k] = \boldsymbol{E}\left[\left(\sqrt{\frac{r}{Y}}\right)^k \cdot \boldsymbol{X}^k\right] = r^{k/2} \cdot \boldsymbol{E}[\boldsymbol{X}^k] \cdot \boldsymbol{E}[{Y}^{-k/2}],
\]
where the expectation is taken componentwise.
\end{proof}

Theorem \ref{th:kst} expresses the $k^{th}$ moment of the NCST random variable as the product of two known components: the moment of the corresponding multivariate skew-normal distribution and the inverse moment of the chi-squared distribution. The following lemma provides a closed-form expression for the latter. 

\begin{lemma}
Let $Y\sim \chi_r^2$. Then for any $k>0$ such that $r>k$, the $k^{th}$ moment of $Y^{-1/2}$ is given by:
\begin{equation}
 \boldsymbol{E}[Y^{-k/2}] = \frac{\Gamma\left(\frac{r-k}{2}\right)}{2^{k/2-1} \cdot \Gamma\left(\frac{r}{2}\right)}.
\end{equation}
\end{lemma}

\begin{proof}
    Let $Y \sim \chi_r^2$. The p.d.f of $Y$ is:
    \[
    f_Y(y) = \frac{1}{2^{r/2} \Gamma(r/2)} y^{r/2-1}e^{-y/2}, \quad y>0.
    \]
    We want to compute the $k$-th moment of $Y^{-1/2}$, i.e., $\boldsymbol{E}[Y^{-k/2}]$:
    \[
    \boldsymbol{E}[Y^{-k/2}] = \int_0^\infty y^{-k/2} f_Y(y) \, dy = \frac{1}{2^{r/2} \Gamma(r/2)} \int_0^\infty y^{(r-k)/2-1} e^{-y/2} \, dy.
    \]
    Let $u = y/2$, so $y=2u$, $dy=2du$. Substituting gives:
    \[
    \boldsymbol{E}[Y^{-k/2}] = \frac{1}{2^{r/2} \Gamma(r/2)} \int_0^\infty (2u)^{(r-k)/2-1} e^{-u} \cdot 2 \, du = \frac{2^{(r-k)/2+1}}{2^{r/2} \Gamma(r/2)} \int_0^\infty u^{(r-k)/2-1} e^{-u} \, du.
    \]
    Simplify: 
    \[
    \boldsymbol{E}[Y^{-k/2}] = \frac{2^{1-k/2}}{\Gamma(r/2)}\cdot \Gamma\left(\frac{r-k}{2}\right) = \frac{\Gamma\left(\frac{r-k}{2}\right)}{2^{k/2-1}\Gamma(r/2)}.
    \]
\end{proof}

% \begin{proof}
% \begin{align}
%  \boldsymbol{E}Y^{-k/2} &= \int_{0}^{ \infty} \quad \! y^{-k/2}~\frac{y^{(\frac{r}{2}-1)} e^{-y/2}~}
% 							{\Gamma(r/2) 2^{r/2}} \, \mathrm{d} y \nonumber\\
% 						&=\frac{1}{\Gamma(r/2) 2^{r/2}} \int_{0}^{ \infty} \quad \! y^{(\frac{r-k}{2}-1)} e^{-y/2}
% 						 \, \mathrm{d} y \nonumber\\
% 						&= \frac{1}{\Gamma(r/2) 2^{r/2}}~\large\Gamma\left(\frac{r-k}{2}\right) 2^{(r-k)/2} , \quad r > k\nonumber\\
% 						&= \frac{\Gamma(\frac{r-k}{2})}{\Gamma(\frac{r}{2})2^{k/2}}. \nonumber
% \end{align}
% \end{proof}

Analytical expressions of Azzalini's multivariate skew-normal distribution are available in the literature \citep[p.534]{WANG2009533}. In particular, the moment-generating function of $\boldsymbol{X} \sim \mathrm{SN}_k(\boldsymbol{\xi}, \boldsymbol{\Omega}, \boldsymbol{\alpha})$ is given by

\[
M_X(\boldsymbol{t}) = 2 \exp{\left(\boldsymbol{t}^\top \xi + \frac{\boldsymbol{t}^\top \boldsymbol{\Omega}\boldsymbol{t}}{2}\right)}\Phi\left(\frac{\boldsymbol{\alpha}^\top \boldsymbol{\Omega} \boldsymbol{t}}{1+\alpha^\top \boldsymbol{\Omega} \boldsymbol{\alpha}}\right),\quad \text{for}~ t\in \mathbb{R}^k,
\]

and the expectation of $\boldsymbol{X}$ is 
\[
\boldsymbol{E}[\boldsymbol{X}]=\boldsymbol{\xi} +\sqrt{\frac{2}{\pi}} \cdot \frac{\boldsymbol{\Omega}\boldsymbol{\alpha}}{\sqrt{{1+\boldsymbol{\alpha}^\top \boldsymbol{\Omega} \boldsymbol{\alpha}}}}.
\]

Applying Theorem~\ref{th:kst}, we obtain the componentwise expectation of $\boldsymbol{T} \sim \mathrm{NCST}_k(\boldsymbol{\xi}, \boldsymbol{\Omega}, \boldsymbol{\alpha}, r)$:
\begin{equation}
\boldsymbol{E}[\boldsymbol{T}]=\sqrt{\frac{r}{2}} \cdot \frac{\Gamma\left(\frac{r-1}{2}\right)}{\Gamma\left(\frac{r}{2}\right)} \cdot \left[\boldsymbol{\xi} +\sqrt{\frac{2}{\pi}} \cdot \frac{\boldsymbol{\Omega}\boldsymbol{\alpha}}{\sqrt{1+\boldsymbol{\alpha}^\top \boldsymbol{\Omega} \boldsymbol{\alpha}}}\right],\quad \text{for}~ r > 1.\\
\end{equation}

Note that $\boldsymbol{E}[\boldsymbol{T}]$ exists only when $r>1$. Higher-order moments can be derived analogously by combining Theorem~\ref{th:kst} with known expressions for higher-order moments of the skew-normal distributions.

\subsection{Affine Transformations}
% \begin{theorem} \label{th:affineMV} \textbf{(The Affine Transformation of the Skew Normal Family):}~\\%omitted , \boldsymbol{\tau}
% Let $\boldsymbol{Z}\sim SN_{d} (\boldsymbol{\xi}, \boldsymbol{\Omega}, \boldsymbol{\alpha}), \boldsymbol{a}\in \boldsymbol{R}^{h} $ and let  $\boldsymbol{B}$ be an $h \times d$ matrix with $h \le d.$
% If $\boldsymbol{W} = \boldsymbol{a} + \boldsymbol{BZ}$, then $\boldsymbol{W}\sim SN_{h}(\boldsymbol{\xi}_{W}, \boldsymbol{\Omega}_{W}, \boldsymbol{\alpha}_{W})$, where
% \begin{equation}
% \begin{split}
% \boldsymbol{\xi}_{W} &= \boldsymbol{a} +\boldsymbol{B\xi},\\
% \boldsymbol{\Omega}_{W}&= \boldsymbol{B\Omega B}^{T} ,\\
% \boldsymbol{\alpha}_{W} &= \frac{1}{\left(1+\boldsymbol{\alpha}^{T}(\bar{\boldsymbol{\Omega}}- \boldsymbol{H\Omega_{W}^{-1}H^{T}}) \boldsymbol{\alpha}\right)^{1/2}}
% \boldsymbol{\omega}_{W}\boldsymbol{\Omega}_{W}^{-1}\boldsymbol{H}^{T}\boldsymbol{\alpha}, \qquad \boldsymbol{H}= \boldsymbol{\omega}^{-1}\boldsymbol{\Omega B}^{T},\\
% \end{split}
% \end{equation}
% $\boldsymbol{\omega}_{W}$ denotes the diagonal matrix of standard deviations of $\boldsymbol{\Omega}_{W},$
% and $\bar{\Omega}= \boldsymbol{\omega}^{-1}\boldsymbol{\Omega} \boldsymbol{\omega}^{-1}$ denotes the correlation matrix associated with $\boldsymbol{\Omega}$.
% \end{theorem}
% \begin{proof}
% See Azzalini (2005) for the proof and more details.
% \end{proof}

A desirable property of multivariate distributions is closure under affine transformations, which facilitates marginal analysis, projection, and model reparameterization. The following theorem shows that the NCST distribution preserves its form under linear transformations of the random vector. 

\begin{theorem}
Let $\boldsymbol{T} \sim \mathrm{NCST}_{k}(\boldsymbol{\xi}, \boldsymbol{\Omega}, \boldsymbol{\alpha}, r)$, ~$\boldsymbol{\xi, \alpha}\in \boldsymbol{R}^{k}$ and let $\boldsymbol{A} \in \mathbb{R}^{k\times h}$ with $h \le k$. Define the linear transformation $\boldsymbol{W} =  \boldsymbol{A^\top T}$. Then $\boldsymbol{W} \sim \mathrm{NCST}_{h}(\boldsymbol{\xi}_{W}, \boldsymbol{\Omega}_{W}, \boldsymbol{\alpha}_{W}, r)$, where
\begin{equation}
\begin{split}
\boldsymbol{\xi}_{W} &= \boldsymbol{A^\top\xi},\\
\boldsymbol{\Omega}_{W}&= \boldsymbol{A^\top \Omega A} ,\\
\boldsymbol{\alpha}_{W} &= \frac{\boldsymbol{\omega}_{W}\boldsymbol{\Omega}_{W}^{-1}\boldsymbol{B}^\top\boldsymbol{\alpha}}{\{1+\boldsymbol{\alpha}^\top ({\boldsymbol{\Omega}_Z}- \boldsymbol{B\Omega}_{W}^{-1}\boldsymbol{B}^\top) \boldsymbol{\alpha}\}^{1/2}}
\end{split}
\end{equation}
with 
\begin{itemize}
    \item $\boldsymbol{B}= \boldsymbol{\omega}^{-1}\boldsymbol{\Omega A}^{T}$,
    \item $\omega =$ the diagonal matrix of the square root of the diagonal elements of $\boldsymbol{\Omega}$,
    \item $\omega_w =$ the diagonal matrix of the square root of the diagonal elements of $\boldsymbol{\Omega_W}$,
    \item $\boldsymbol{\Omega}_Z=\boldsymbol{\omega}^{-1}\Omega \omega^{-1}$ (the standardized covariance of $\boldsymbol{Z} \sim \mathrm{SN}_k(\boldsymbol{\Omega}_Z, \boldsymbol{\alpha})$) 
\end{itemize}
\end{theorem}

\begin{proof}
Let $\boldsymbol{T} = \frac{\boldsymbol{X}}{\sqrt{Y/r}}$, where
\begin{itemize}
    \item $\boldsymbol{X} \sim \mathrm{SN}_k(\boldsymbol{\xi}, \boldsymbol{\Omega}, \boldsymbol{\alpha})$, 
    \item $Y \sim \chi_r^2$, independent of $\boldsymbol{X}$. 
\end{itemize}

By the linearity of affine transformations and independence from the scaling $\sqrt{Y/r}$, we consider:
\[
\boldsymbol{W} = \boldsymbol{A}^\top \boldsymbol{T} = \frac{\boldsymbol{A}^\top \boldsymbol{X}}{\sqrt{Y/r}}.
\]

Now, $\boldsymbol{X}_W:= \boldsymbol{A}^\top \boldsymbol{X} \sim \mathrm{SN}_h(\boldsymbol{\xi}_W, \boldsymbol{\Omega}_W, \boldsymbol{\alpha}_W)$ based on the affine transformation property of skew-normal variables from \cite[p.584-585]{Azzalini1999}, using the parameters given above. 

Hence, since $\boldsymbol{W} = \frac{\boldsymbol{X}_{W}}{\sqrt{Y/r}}$, it follows that $\boldsymbol{W} \sim \mathrm{NCST}_h(\boldsymbol{\xi}_{W}, \boldsymbol{\Omega}_{W}, \boldsymbol{\alpha}_{W}, r)$. 

\end{proof}

%The location-scale extension of Proposition 5 in \cite{Azzalini1999} implies that if $\boldsymbol{X}\sim SN_{k} (\boldsymbol{\xi}, \boldsymbol{\Omega}, \boldsymbol{\alpha})$, then $\boldsymbol{A^\top X}\sim SN_{k}(\boldsymbol{\xi}_{W}, \boldsymbol{\Omega}_{W}, \boldsymbol{\alpha}_{W})$. Definition (\ref{de:MNST}) implies that if $Y$ is a univariate $\chi^2$ random variable with $r$ degrees of freedom which is independent of $\boldsymbol{X}$,then $$\frac{\boldsymbol{A^\top X}}{\sqrt{Y/r}}\sim NCST_{h}(\boldsymbol{\xi}_{W}, \boldsymbol{\Omega}_{W}, \boldsymbol{\alpha}_{W}, r). $$ Therefore, $$\boldsymbol{W}\sim NCST_{h}(\boldsymbol{\xi}_{W}, \boldsymbol{\Omega}_{W}, \boldsymbol{\alpha}_{W}, r). $$

\subsection{Quadratic Forms and Their Distribution}\label{sec:quadratic}

Distributions of quadratic forms involving skew-normal and skew $t$ variables have been investigated in various studies due to their relevance in statistical inference. In particular, understanding how these quadratic forms behave under skewed distributions motivates the development of generalized versions of classical results such as the chi-squared and $F$ distributions. 

To set the stage for our results, we begin by reviewing classical and skewed extensions of the chi-squared and $F$ distributions. These definitions provide the necessary background for the theorems that follow. 

\begin{definition}[Non-central Chi-squared Distribution]\label{de:MNST}
Let $X_1, X_2, \dots, X_n$ be independent random variables such that $X_i \sim \mathrm{N}(\mu_i, \sigma^2)$ for $i = 1, \dots, n$. The random variable
\[
Y = \frac{1}{\sigma^2} \sum_{i=1}^{n} X_i^2
\]
follows a non-central chi-squared distribution with $n$ degrees of freedom and non-centrality parameter
\[
\delta = \frac{1}{\sigma^2} \sum_{i=1}^{n} \mu_i^2.
\]

When $\delta = 0$, the distribution reduces to the \textit{central} chi-square distribution.
\end{definition}

Building on this, \cite{WANG2009533} introduced a skewed version of the non-central chi-squared distribution by replacing the normal distribution with the multivariate skew-normal distribution. 

\begin{definition}[Non-central Skew Chi-squared Distribution]\label{de:schi}
Let $\boldsymbol{X}\sim \mathrm{SN}_{k}(\boldsymbol{\xi} ,\boldsymbol{ I}_{k}, \boldsymbol{\alpha} )$. Then the distribution of $Q = \boldsymbol{X}^\top \boldsymbol{X}$ is called the non-central skew-chi-squared distribution with degrees of freedom $k$, non-centrality parameter $\lambda = \boldsymbol{\xi}^\top \boldsymbol{\xi}$, and skewness parameter $\boldsymbol{\alpha}$, denoted: \[
Q \sim S\chi_k^{2}(\lambda, \boldsymbol{\alpha}).
\] 
When $\boldsymbol{\alpha} = \mathbf{0}$, this reduces to the usual non-central chi-squared distribution. 
\end{definition}

Similarly, the classical $F$ distribution has been extended to incorporate non-centrality and skewness in the numerator or denominator. These generalizations, introduced by \cite{johnson1995continuous}, are essential for characterizing the distribution of the quadratic forms involving NCST variables.

\begin{definition}[Doubly non-central $F$ Distribution]\label{de:ncf}
 Let $X_1 \sim \chi^2_{\nu_1}(\lambda_1)$ and $X_2 \sim \chi^2_{\nu_2}(\lambda_2)$ be independent. Then the random variable
 \[
 F = \frac{X_1/\nu_1}{X_2/\nu_2}
 \]
 is said to follow a doubly non-central $F$ distribution with degrees of freedom $(\nu_1, \nu_2)$ and non-centrality parameters $(\lambda_1, \lambda_2)$, denoted by 
 \begin{equation}\label{eq:ncf}
      F \sim F''_{\nu_1, \nu_2}(\lambda_1, \lambda_2).
 \end{equation}
 If $\lambda_{2}= 0$ , the distribution is referred to as a singly non-central skew $F$ distribution and is denoted by $F^{'}_{\nu_{1}, \nu_{2}}(\lambda_{1})$.
\end{definition}

We now define a skewed extension of this distribution, which naturally arises when the numerator follows a skewed distribution. 

\begin{definition}[Non-central Skew $F$ Distribution]\label{de:sncf}
Let $X \sim S\chi_{\nu_{1}}^2(\lambda_{1}, \alpha )$.
Let  $Y\sim \chi_{\nu_{2}}^2(\lambda_{2})$ be independent of $X.$  Then random variable
\begin{equation}
F = \frac{X/ \nu_{1}}{Y/ \nu_{2}},
\end{equation}
is said to follow a doubly non-central skew $F$ distribution with degrees of freedom $(\nu_{1}, \nu_{2})$, non-centrality parameters $\lambda_{1}, \lambda_{2}$ and skewness coefficient $\alpha$, denoted by 
\[
SF^{''}_{\nu_{1}, \nu_{2}}(\lambda_{1} ,\lambda_{2}, \alpha).
\]
If $\lambda_{2}= 0$, the distribution is referred to as a singly non-central skew $F$ distribution and is denoted by $SF^{'}_{\nu_{1}, \nu_{2}}(\lambda_{1}, \alpha)$.
\end{definition}

With these definitions in place, we now derive the distribution of quadratic forms involving the NCST distributions.

\begin{theorem}
Let $\boldsymbol{T}\sim \mathrm{NCST}_{k}(\boldsymbol{\xi}, \boldsymbol{I}_{k}, \boldsymbol{\alpha}, r)$. Then the quadratic form $Q = \boldsymbol{T}^\top \boldsymbol{T}$ follows a singly non-central skew $F$ distribution with degrees of freedom $(1,r)$, non-centrality parameter $\lambda = \boldsymbol{\xi}^{\top}\boldsymbol{\xi}$, and the skewness parameter $\boldsymbol{\alpha}$, denoted:
\[
Q \sim SF'_{1,r}(\lambda, \boldsymbol{\alpha}).
\]
\end{theorem}
\begin{proof}
Let $\boldsymbol{T} = \boldsymbol{X}/\sqrt{Y/r}$, where $\boldsymbol{X} \sim \mathrm{SN}_k(\boldsymbol{\xi}, \boldsymbol{I}_k, \boldsymbol{\alpha})$, and $Y \sim \chi_r^2$ be independent of $\boldsymbol{X}$. Then 
\[
Q = \boldsymbol{T}^\top \boldsymbol{T} = \frac{\boldsymbol{X}^\top \boldsymbol{X}}{Y/r},
\]
where $\boldsymbol{X}^\top \boldsymbol{X} \sim S\chi_k^{'2}(\lambda, \boldsymbol{\alpha})$ as defined in Definition \ref{de:schi}. Therefore, $Q$ is a ratio of independent random variables, with a skewed chi-squared numerator and a chi-squared denominator, which defines a skew $F$ distribution as in Definition \ref{de:sncf}.
That is, the distribution of $Q$ is the singly non-central skew $F$ distribution
with degrees of freedom $(1,r)$, non-centrality parameters $\lambda$ and skewness coefficient $\boldsymbol{\alpha}$, denoted by $Q \sim SF^{'}_{1,r}(\lambda, \boldsymbol{\alpha})$.
\end{proof}

We now generalize this result to arbitrary positive definite scale matrices $\boldsymbol{\Omega}$ and to general quadratic forms. The following theorem establishes necessary and sufficient conditions under which a quadratic form of an NCST distributed random vector follows a skew $F$ distribution. 

\begin{theorem}[Quadratic Forms of NCST Distributions]\label{thm:quadratic} 
    Let $\boldsymbol{T} \sim \mathrm{NCST}_k(\boldsymbol{\xi}, \boldsymbol{\Omega}, \boldsymbol{\alpha}, r)$, and define $Q = \boldsymbol{T}^\top \boldsymbol{W} \boldsymbol{T}$ for a symmetric, non-negative definite matrix $\boldsymbol{W} \in \mathbb{R}^{k\times k}$. Then the necessary and sufficient conditions under which 
    \[
    Q \sim SF^{'}_{1, r}(\lambda, \alpha_{*}),
    \]
     for some $\boldsymbol{\alpha_{*}}$ including $\boldsymbol{\alpha_{*}}=\boldsymbol{0}$, are:
     \begin{enumerate}
        \item[(i)] $\boldsymbol{\Omega}^{1/2} W \boldsymbol{\Omega}^{1/2}$  is idempotent with rank $m = \mbox{rank}(\boldsymbol{\Omega}^{1/2} W \boldsymbol{\Omega}^{1/2})$, 
        \item[(ii)] $\lambda =\boldsymbol{\xi}^{\top} \boldsymbol{W \xi} = \boldsymbol{\xi}^{\top} \boldsymbol{W}^{\top} \boldsymbol{\Omega}\boldsymbol{W \xi}$,
        \item[(iii)] $\boldsymbol{\alpha}^{\top} \boldsymbol{\Omega}^{1/2}\boldsymbol{W \xi} = c_\alpha \cdot \boldsymbol{\alpha}_{*}^{\top} \boldsymbol{\nu}$,
        \item[(iv)] $\boldsymbol{\alpha}_{*}^{\top}\boldsymbol{\alpha}_{*} = c_\alpha^{-2} \boldsymbol{\alpha}^{\top} \boldsymbol{P}_{1}\boldsymbol{P}_{1}^{\top} \boldsymbol{\alpha}$,\\
        where
        \begin{itemize}
            \item $c_\alpha = \sqrt{1+ \boldsymbol{\alpha}^{\top}\boldsymbol{P}_{1}\boldsymbol{P}_{1}^{\top} \boldsymbol{\alpha}}$, 
            \item $ \boldsymbol{\nu}= \boldsymbol{P}_{1}^{\top}\boldsymbol{W}\boldsymbol{\Omega}^{1/2}\boldsymbol{\xi}$,
            \item $\boldsymbol{P}=(\boldsymbol{P}_{1}, \boldsymbol{P}_{2})$ is an orthogonal matrix in $\mathbb{R}^{ k\times k}$ such that
        \begin{equation}
        \boldsymbol{\Omega}^{1/2}\boldsymbol{W}\boldsymbol{\Omega}^{1/2} = \boldsymbol{P} \left(
        \begin{matrix}
        I_{m} & 0 \\
        0  & 0
        \end{matrix}
         \right) \boldsymbol{P}^{\top} = \boldsymbol{P}_{1} \boldsymbol{P}_1^{\top}.
        \end{equation}
        \end{itemize}
\end{enumerate}
\end{theorem}

\begin{proof}
 Let $\boldsymbol{T} = \frac{\boldsymbol{X}}{\sqrt{Y/r}}$, with $\boldsymbol{X} \sim \mathrm{SN}_k(\boldsymbol{\xi}, \boldsymbol{\Omega}, \boldsymbol{\alpha})$, and $Y \sim \chi_r^2$ be independent of $\boldsymbol{X}$. Then the quadratic form becomes:
\[
Q = \boldsymbol{T}^\top \boldsymbol{W} \boldsymbol{T} = \frac{\boldsymbol{X}^\top \boldsymbol{W} \boldsymbol{X}}{Y/r}.
\]
By Theorem 4.1 in \citet[p.540]{WANG2009533}, the conditions (i)-(iv) are necessary and sufficient for $\boldsymbol{X}^\top \boldsymbol{W} \boldsymbol{X} \sim S\chi_m^2(\lambda, \alpha_*)$, where $m = \text{rank}(\boldsymbol{\Omega}^{1/2} W \boldsymbol{\Omega}^{1/2})$.   
Definition~\ref{de:sncf} implies that Since $Y$ is independent of $\boldsymbol{X}^\top \boldsymbol{W} \boldsymbol{X}$ and follows $\chi_r^2$, the distribution of $Q$ is the singly non-central skew $F$ distribution with degrees of freedom $(1,r)$, non-centrality parameters $\lambda$ and skewness coefficient $\boldsymbol{\alpha}$, denoted by $Q \sim SF^{'}_{1,r}(\lambda, \boldsymbol{\alpha})$.
\end{proof}

\section{Simulation Study}\label{sec:simulation}
To complement the theoretical development of the multivariate NCST distribution, we conduct a simulation study to empirically examine its distributional characteristics. While the proposed distribution is defined in an arbitrary dimension $k$, we focus on the bivariate case ($k=2$) as an example to illustrate the effects of skewness, tail behavior, and properties of the quadratic form. The simulation is designed to validate the theoretical results presented in the earlier sections and to visualize how key distributional parameters influence marginal and joint behavior. 

The simulation is based on the stochastic representation of the multivariate NCST distribution described in Definition \ref{de:NCST}. Specifically, we generate random vectors from $\mathrm{NCST}_{k}(\boldsymbol{\xi},\boldsymbol{\Omega}, \boldsymbol{\alpha}, r)$ by applying the transformation in equation (\ref{eq:mvst1}) to samples from the multivariate skew-normal distribution with independent chi-squared noise. Unless otherwise stated, we fix the location and scale parameters as $\boldsymbol{\xi} = (1, 2)^\top$ and $\boldsymbol{\Omega} = diag(4, 1)$, and consider a range values for the skewness vector $\boldsymbol{\alpha}$ and degrees of freedom $r$. For each simulation scenario, we generate $n=100,000$ independent samples from the bivariate NCST distribution. 

\subsection{Quadratic Form Validation}

One of the key theoretical results established is Theorem \ref{thm:quadratic}, which characterizes the distribution of quadratic forms of multivariate NCST variables. Specifically, if~$\boldsymbol{T} \sim \mathrm{NCST}_{k}(\boldsymbol{\xi},\boldsymbol{\Omega}, \boldsymbol{\alpha}, r)$ and $Q = \boldsymbol{T}^\top \boldsymbol{W} \boldsymbol{T}$ for a symmetric, nonnegative definite matrix $\boldsymbol{W}$, then $Q$ follows a non-central skew $F$ distribution under certain algebraic conditions. 

To validate Theorem \ref{thm:quadratic}, we designed a simulation study in which the conditions of the theorem are explicitly satisfied. We set $\boldsymbol{\Omega} = \boldsymbol{I}_2$ and constructed the projection matrix $\boldsymbol{W} = \boldsymbol{a}\boldsymbol{a}^\top/\|a\|^2$ with $\boldsymbol{a} = (1,1)^\top$, so that $\boldsymbol{\Omega}^{1/2} W \boldsymbol{\Omega}^{1/2}=\boldsymbol{W}$ is idempotent and of rank one. Under this setup, the key parameters of the resulting non-central skew $F$ distribution can be computed as follows:

\[
\lambda = \boldsymbol{\xi}^\top \boldsymbol{W} \boldsymbol{\xi} = 4.5, \quad
\boldsymbol{P} = \begin{pmatrix}
\frac{1}{\sqrt{2}} & \frac{1}{\sqrt{2}} \\
-\frac{1}{\sqrt{2}} & \frac{1}{\sqrt{2}}
\end{pmatrix}, \quad
c = \sqrt{19}, \quad \nu = \frac{3}{\sqrt{2}}, \quad \alpha_\star = \sqrt{\frac{18}{19}}.
\]
Thus, the quadratic form $Q = \boldsymbol{T}^\top \boldsymbol{W} \boldsymbol{T}$ satisfies the conditions for which $Q \sim SF_{1,3}'(4.5, \sqrt{18/19})$.

We generated random samples from $\mathrm{NCST}_{2}((1,2)^\top, \boldsymbol{I}_2, (3,3)^\top, 3)$, and computed the corresponding quadratic forms $\boldsymbol{Q} = \boldsymbol{T}^\top \boldsymbol{WT}$. Since the exact density function of the non-central skew $F$ distribution is not analytically tractable and no R package provides a closed-form implementation or random variate generator for this distribution, we instead constructed a reference distribution via simulation. Specifically, we generated non-central skew $F$ variables by $F = \frac{Z^2/1}{Y/3}$, where $Z \sim \mathrm{SN}(\sqrt{\lambda}, 1, \alpha_*)$, where $\lambda=4.5$ and $\alpha_*=\sqrt{18/19}$ and $Y \sim \chi^2_3$ independently. This simulated reference mimics the defining structure of the non-central skew $F$ distribution $SF_{1,3}'(4.5, \sqrt{18/19})$. 

\begin{figure}[ht]
    \centering
    \includegraphics[width=0.87\linewidth]{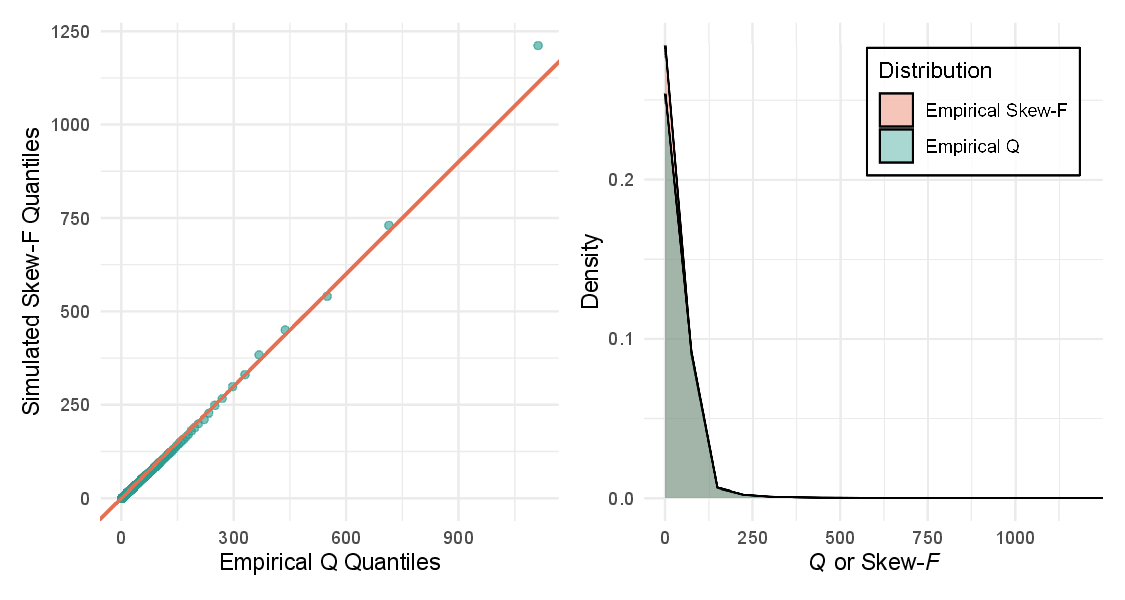}
    \caption{Left: Q-Q plot comparing the empirical distribution of $Q$ with a simulated non-central skew $F$ distribution. The 100th percentile was omitted to reduce distortion from extreme tail behavior. Right: Overlaid kernel density estimates of the same two distributions, truncated at the 99.9th percentile.}
    \label{fig:sim_Q}
\end{figure}

To assess agreement between the two distributions, we compared the empirical distribution of $Q$ with that of the simulated skew $F$ variable using a quantile-quantile plot and overlaid kernel density estimates, as shown in Figure \ref{fig:sim_Q}. To improve visual interpretability and reduce the influence of extreme tail values, both plots were based on the first 99.9th percentiles of the data. The results show close agreement across most quantiles and nearly indistinguishable density shapes aside from minor deviations in the tails, which provides empirical support for the theoretical distributional result in Theorem \ref{thm:quadratic}.

\subsection{Skewness and Tail Behavior}
To examine how the skewness parameter $\boldsymbol{\alpha}$ influences the distributional shape of the bivariate NCST distribution. We conduct a simulation study under three skewness levels: $\boldsymbol{\alpha} = (0,0)^\top$ (symmetric), $\boldsymbol{\alpha} = (3, 3)^\top$ (moderate skew), and $\boldsymbol{\alpha} = (15,15)^\top$ (severe skew). We fix the degrees of freedom, $r=3$, and use equal skewness in both dimensions to isolate the effect of skewness intensity while maintaining a visually balanced joint distribution. 

\begin{figure}[ht]
    \centering
    \includegraphics[width=0.87\linewidth]{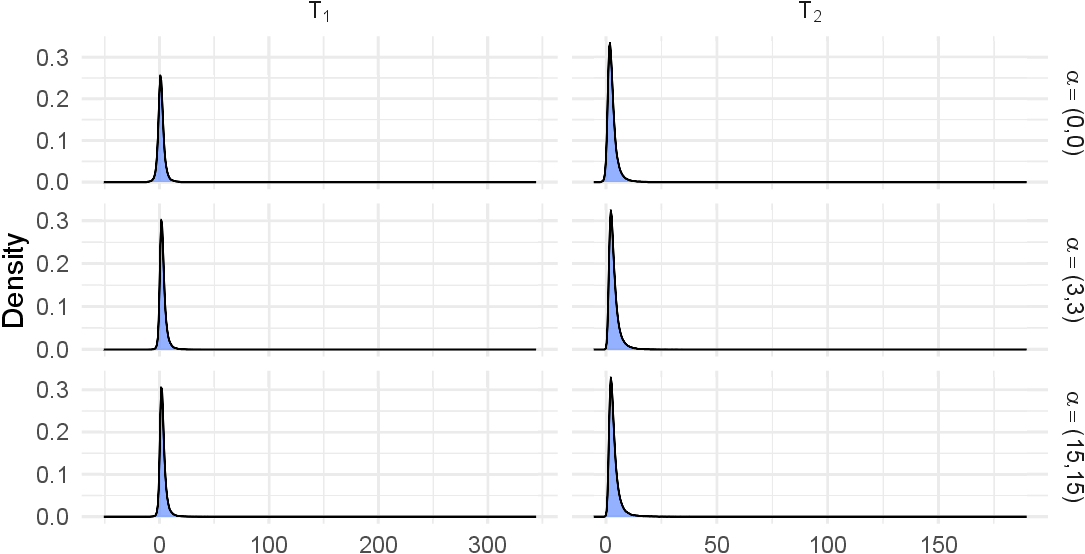}
    \caption{Marginal density plots of $T_1$ and $T_2$ under varying $\boldsymbol{\alpha^\top} = (0, 0),\ (3, 3),\ (15, 15)$.}
    \label{fig:sim_alpha_marginals}
\end{figure}

Figure \ref{fig:sim_alpha_marginals} displays the marginal densities of $T_1$ and $T_2$ across the three skewness settings. Each panel is based on $n = 100{,}000$ simulated observations from the bivariate NCST distribution with fixed location $\boldsymbol{\xi} = (1, 2)^\top$ and scale $\boldsymbol{\Omega} = diag(4, 1)$. As skewness increases, the marginal distributions become more sharply peaked and exhibit heavier right tails, particularly in $T_1$. This is visually reflected in the narrowing and rightward stretching of the blue density regions. These effects are more noticeable in $T_1$ than $T_2$, likely due to its larger scale parameter $\sigma_1=2$, which amplifies the influences of the skewness term. 

To quantify these differences, Table \ref{tab:skewness_quantiles} reports the empirical skewness and 95th percentiles for $T_1$ and $T_2$ across the three cases. The skewness values increase significantly with higher $\alpha$, especially in $T_1$, where skewness increases from 2.79 to 15.40. The 95th percentiles are also shifted upward, reflecting heavier right tails. Interestingly, the increase in the 95th percentiles for both $T_1$ and $T_2$ begins to taper between the moderate and severe skewness cases, suggesting a saturation effect at extreme levels of skewness. This may reflect the increasingly heavy right tails where additional skew diminishes marginal influence on the upper quantiles. Although only three settings are presented here for clarity, similar trends were observed across a broader range of $\boldsymbol{\alpha}$ values examined during simulation. 

Although the empirical skewness increases with $||\boldsymbol{\alpha}||$, its magnitude may vary significantly across simulation replications, especially in high skewness settings. In contrast, quantile-based tail measures such as the 95th percentile show more stable trends and may serve as more robust indicators of tail heaviness in practice. 

\begin{table}[ht]
\centering
\caption{Empirical skewness and 95th percentiles for $T_1$ and $T_2$ under varying $\boldsymbol{\alpha}$.}
\begin{tabular}{lcccc}
\toprule
\textbf{$\boldsymbol{\alpha}$} & Skew$(T_1)$ & Skew$(T_2)$ & 95th Percentile $(T_1)$ & 95th Percentile $(T_2)$ \\
\midrule
$(0,\ 0)$   & 2.79 & 6.33 & 6.70 & 6.83 \\
$(3,\ 3)$   & 4.53 & 5.58 & 8.42 & 8.03 \\
$(15,\ 15)$ & 15.40 & 9.96 & 8.33 & 7.98 \\
\bottomrule
\end{tabular}
\label{tab:skewness_quantiles}
\end{table}

\begin{figure}[ht]
    \centering
    \includegraphics[width=0.9\linewidth]{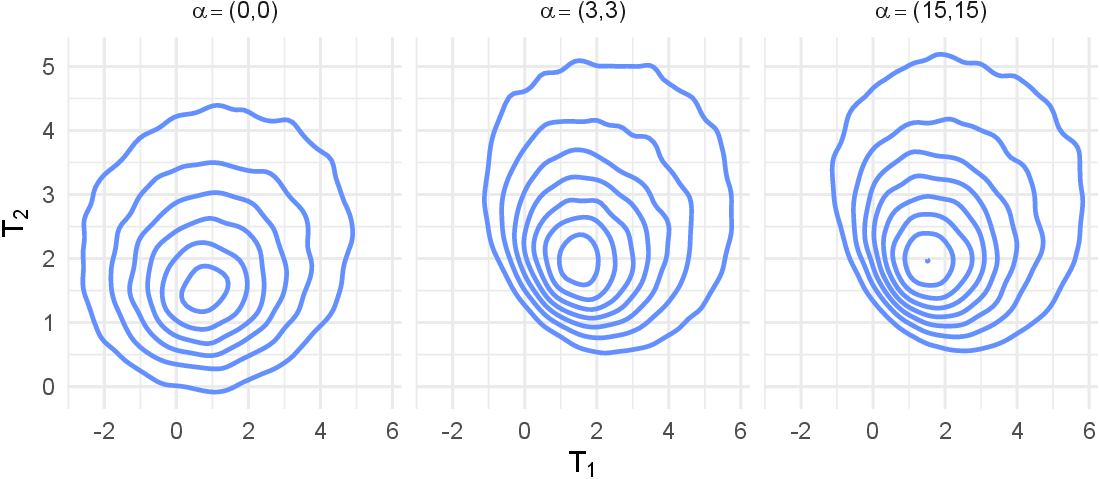}
    \caption{Contour Plots of Joint Density under varying $\boldsymbol{\alpha}$.}
    \label{fig:sim_alpha_contours}
\end{figure}

Figure \ref{fig:sim_alpha_contours} shows the joint density contour plots for each skewness level. The contours become increasingly stretched and skewed in the direction of the positive orthant as $\boldsymbol{\alpha}$ increases. In the symmetric case, the density is nearly elliptical, while for $\boldsymbol{\alpha}=(15,15)^\top$, the mass is clearly concentrated along the skew direction, and the density peak shifts upward and rightward. This visualization confirms the directional impact of skewness on the joint distribution, which complements the marginal behavior observed earlier. 

\subsection{Effect of Degrees of Freedom}

This subsection explores how varying the degrees of freedom $r$ affects the shape and tail behavior of the bivariate NCST distribution. As $r$ increases, the distribution is expected to become lighter-tailed and converge to its skew-normal counterpart due to the underlying scale-mixture representation. To illustrate this transition, we fixed the skewness parameter at $\boldsymbol{\alpha} = (3,3)^\top $ and consider four representative values for $ r = 3, 5, 10, 30$. All simulations use the same location and scale parameters as introduced previously. 

\begin{figure}[ht]
    \centering
    \includegraphics[width=0.9\linewidth]{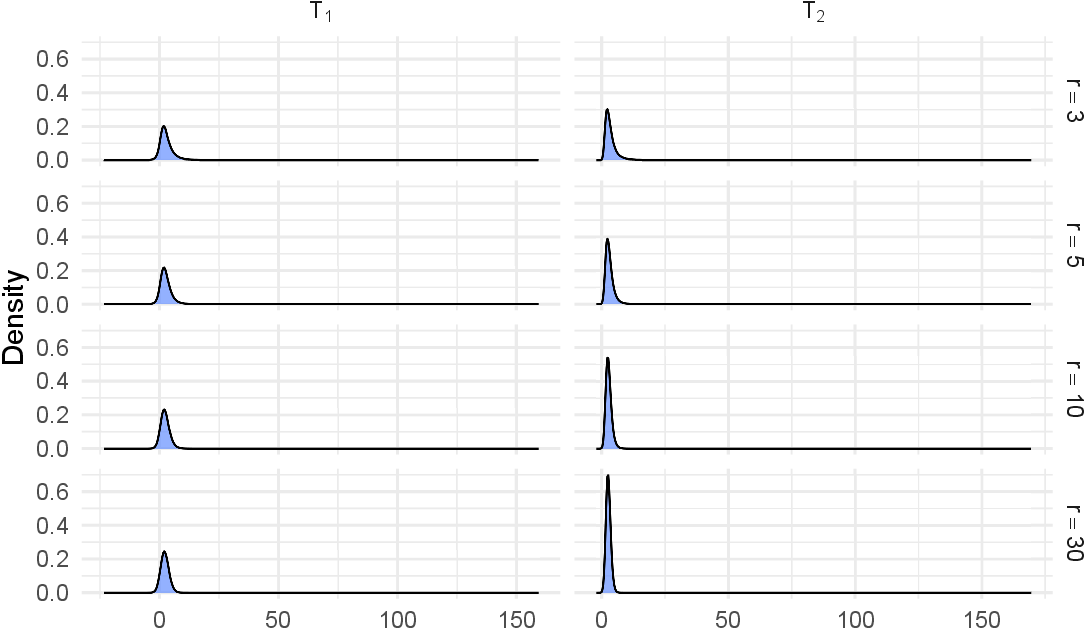}
    \caption{Marginal densities of $T_1$ and $T_2$ across degrees of freedom.}
    \label{fig:sim_densitiy_degrees}
\end{figure}

Following the structure of the previous subsection, we first examine the marginal densities of $T_1$ and $T_2$ under varying degrees of freedom, as shown in Figure \ref{fig:sim_densitiy_degrees}. As $r$ increases, the peak of the marginal distributions becomes sharper and the tail becomes more elongated. At lower values of $r$, the densities are more dispersed and right-skewed, reflecting visibly heavier tails. The shift toward symmetry and concentration becomes especially apparent at $r=30$, where the density closely resembles that of a skew-normal distribution. These trends reinforce the role of $r$ in controlling tail thickness and concentration in the NCST distribution.

Table \ref{tab:df_skewness_summary} reports empirical skewness, (excess) kurtosis, and 95th percentiles for $T_1$ and $T_2$ across the selected values of $r$. As expected, skewness and kurtosis decrease substantially with increasing $r$, reflecting a transition from heavy-tailed and asymmetric distributions to lighter-tailed and more symmetric ones. For example, at $r=3$, both $T_1$ and $T_2$ exhibit extreme kurtosis (121 and 187, respectively), while at $r=30$, the kurtosis values drop below 1. The 95th percentiles follow a similar trend, showing reduced upper-tail mass as $r$ increases. These results align with the theoretical convergence of the NCST to its skew-normal counterpart as $r \to \infty$, due to the underlying scale-mixture representation.

\begin{table}[ht]
\centering
\caption{Empirical skewness, kurtosis, and 95th percentiles for $T_1$ and $T_2$ under varying degrees of freedom $r$.}
\begin{tabular}{lcccccc}
\toprule
\textbf{$r$} & Skew$(T_1)$ & Skew$(T_2)$ & Kurt$(T_1)$ & Kurt$(T_2)$ & 95th \%ile $(T_1)$ & 95th \%ile $(T_2)$ \\
\midrule
3  & 6.02  & 7.41  & 121.00 & 187.00 & 8.41 & 8.08 \\
5  & 1.86  & 2.79  & 21.80  & 29.40  & 6.65 & 6.02 \\
10 & 0.64  & 1.02  & 1.76   & 2.55   & 5.65 & 4.86 \\
30 & 0.25  & 0.40  & 0.36   & 0.41   & 5.15 & 4.25 \\
\bottomrule
\end{tabular}
\label{tab:df_skewness_summary}
\end{table}

\begin{figure}[ht]
    \centering
    \includegraphics[width=0.9\linewidth]{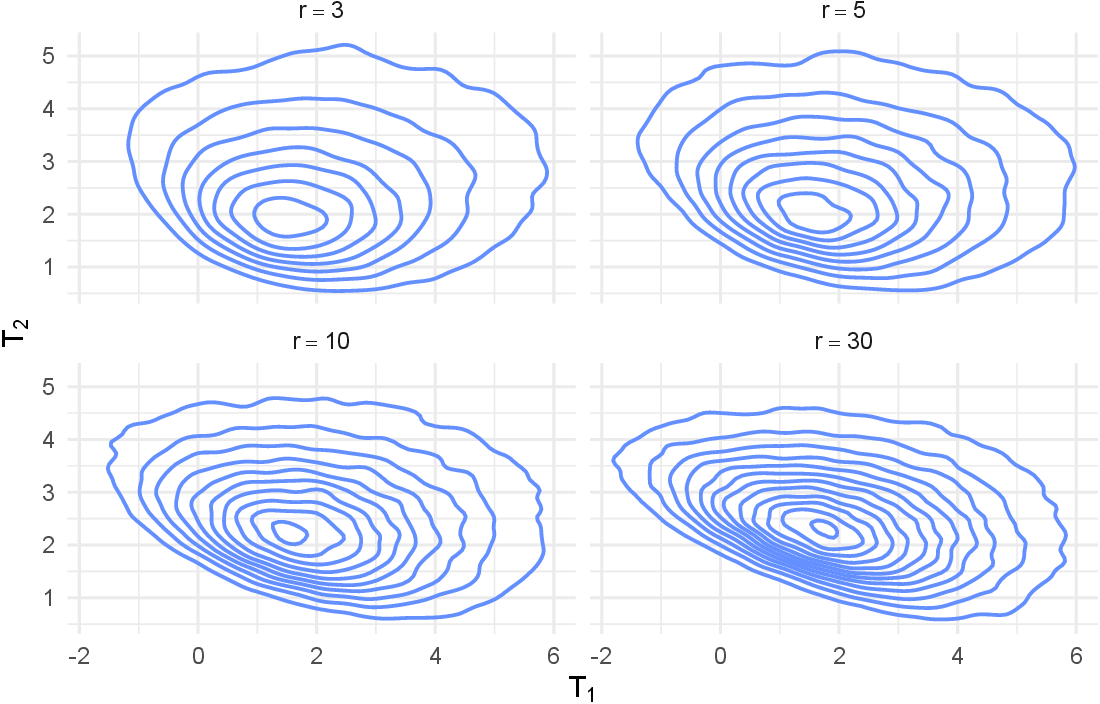}
    \caption{Joint density contours of ($T_1$, $T_2$) under varying degrees of freedom.}
    \label{fig:dim_df_contours}
\end{figure}

Figure \ref{fig:dim_df_contours} displays the joint density contours of $(T_1, T_2)$ for the same set of degrees of freedom. As $r$ increases from 3 to 30, the contours become progressively more symmetric and concentrated around the center, indicating the reduction in tail heaviness and the approach toward a skew-normal shape. At low values of $r$, such as $r=3$ and $r=5$, the contours are elongated and exhibit strong skewness. In contrast, the $r=30$ case shows an "egg-like" shape, nearly elliptical with subtle asymmetry, visually reinforcing the convergence suggested by the marginal densities and summary statistics. 

\subsection{Comparison to Misspecified Models}

We compare the bivariate NCST distribution to three baseline alternatives: the multivariate $t$ distribution, the multivariate normal distribution, and Azzalini's multivariate skewed $t$ distribution \citep{Azzalini_2013}. For this demonstration, we simulate data from an NCST distribution with location parameter $\boldsymbol{\xi} = (1, 2)^\top$, scale matrix $\boldsymbol{\Omega} = \text{diag}(4,1)$, $\boldsymbol{\alpha} = (3,0)^\top$, and degrees of freedom $r=1$, and sample size $n=100$. These settings produce a dataset with strong asymmetry and heavy tails. 

Unlike the Azzalini-type skew $t$ distribution, which admits a closed-form expression, the NCST does not due to its non-central construction. To evaluate the likelihood under the NCST model, we approximate its marginal density via Monte Carlo integration. Specifically, we draw $y_i \sim \chi_r^2$ for $i=1, \dots, M$ and estimate the marginal density at a point $\boldsymbol{t} \in \boldsymbol{R}^k$ as 
\[
\hat{f}(\boldsymbol{t}) = \frac{1}{M} \sum_{i=1}^{M} 
f_{\text{SN}_k}\left(\boldsymbol{t} \cdot \sqrt{y_i / r}; \boldsymbol{\xi}, \boldsymbol{\Omega}, \boldsymbol{\alpha} \right) \cdot 
\left( \sqrt{y_i / r} \right)^k
\]
where the second factor accounts for the Jacobian of the transformation from $\boldsymbol{X}\sim \text{SN}_k(\boldsymbol{\xi}, \boldsymbol{\Omega}, \boldsymbol{\alpha})$ to $\boldsymbol{T} = \frac{\boldsymbol{X}}{\sqrt{Y/r}}$. This estimator arises from marginalizing over the mixing variable in the NCST's stochastic representation. 

We fit each of the four models to the same dataset, maximizing the likelihood using either its closed-form expression (when available) or this simulation-based approximation. Akaike Information Criterion (AIC) and Schwarz Information Criterion (SIC) are computed to compare model fit while penalizing model complexity. As shown in Table \ref{tab:model_comparison}, the NCST model achieves the highest log-likelihood and the lowest AIC and BIC, which indicates a better fit to the data from which it was generated. Notably, Azzalini’s skew-$t$ model effectively captures asymmetry and heavy tails, but the NCST model achieves a higher likelihood with the same model complexity, resulting in lower AIC and SIC values.

\begin{table}[ht]
\centering
\caption{Model comparison based on log-likelihood, AIC, and SIC.}
\label{tab:model_comparison}
\begin{tabular}{lrrr}
\toprule
\textbf{Model} & \textbf{Log-Likelihood} & \textbf{AIC} & \textbf{SIC}\\
\midrule
Multivariate Normal        & $-1051.79$ & $2113.57$ & 2126.60\\
Skew-Normal                & $-984.72$  & $1983.45$ & 2001.68\\
Azzalini’s Skew-$t$        & $-571.39$  & $1158.78$ & 1179.63\\
\textcolor{red}{NCST}      & \textcolor{red}{$-565.95$}  & \textcolor{red}{$1147.89$} & \textcolor{red}{1168.73}\\
\bottomrule
\end{tabular}
\end{table}

\section{Application: Modeling Skewed Tumor Shape Features}\label{sec:applications}
We illustrate the benefits of flexible multivariate distribution modeling using the Breast Cancer Wisconsin Diagnostic Data \citep{breast_cancer_wisconsin_(diagnostic)_17}. The dataset contains diagnostic measurements of breast masses derived from digitized images of fine needle aspirates (FNA), with the aim of distinguishing between malignant and benign tumors. Each of the 569 patient samples is described by 30 continuous features computed from tumor boundary characteristics. 

For this application, we focus on three biologically related variables: \texttt{concavity\_se}, \\ \texttt{symmetry\_se}, and \texttt{fractal\_dimension\_se}. These features quantify variability in the local geometry of the tumor contour and represent standard error measurements, indicating within-sample variability in shape attributes. Such variables are inherently non-negative, often exhibit right-skewed distributions, and may be correlated in asymmetric ways, which makes them well-suited for modeling using multivariate skewed distributions. 

To assess the modeling performance, we fit four candidate models to the joint distribution of these three features: the multivariate normal, multivariate skew-normal, Azzalini's skew-\textit{t}, and our NCST distributions. Similar to the simulation study, the log-likelihood of NCST distributions is approximated via Monte Carlo integration, which may introduce minor variability in likelihood estimates. Model performance is compared using log-likelihood, AIC, and BIC, as summarized in Table \ref{tab:application}. 

\begin{table}[ht]
\centering
\caption{Comparison of four candidate models fitted to the joint distribution of three tumor shape variability features from the Breast Cancer Wisconsin Diagnostic Data.}
\label{tab:application}
\begin{tabular}{lrrr}
\toprule
\textbf{Model} & \textbf{Log-Likelihood} & \textbf{AIC} & \textbf{SIC}\\
\midrule
Multivariate Normal        &  5.64  &   6.72     &  45.81\\
Skew-Normal                & 277.96 & $-531.92$  & $-479.80$\\
Azzalini’s Skew-$t$        & 901.59 & $-1777.17$ & $-1720.70$\\
\textcolor{red}{NCST}      & \textcolor{red}{$940.04$}  & \textcolor{red}{$-1854.09$} & \textcolor{red}{-1797.62}\\
\bottomrule
\end{tabular}
\end{table}

Table \ref{tab:application} shows that the NCST model provided the best overall fit to the data, with the highest log-likelihood (940.04), the lowest AIC ($-1854.09$), and BIC ($-1797.62$) among all models considered. Both versions of the skew \textit{t} distribution substantially outperformed the multivariate normal and skew-normal models, which yielded significantly higher AIC and BIC values. These results highlight the importance of accounting for both skewness and heavy tails when modeling tumor shape variability features. The performance gap between the NCST model and Azzalini's skew-\textit{t} suggests that added flexibility in the non-central construction can further enhance goodness-of-fit, even when evaluated under approximate likelihoods. 

\begin{figure}[ht]
    \centering
    \includegraphics[width=\linewidth]{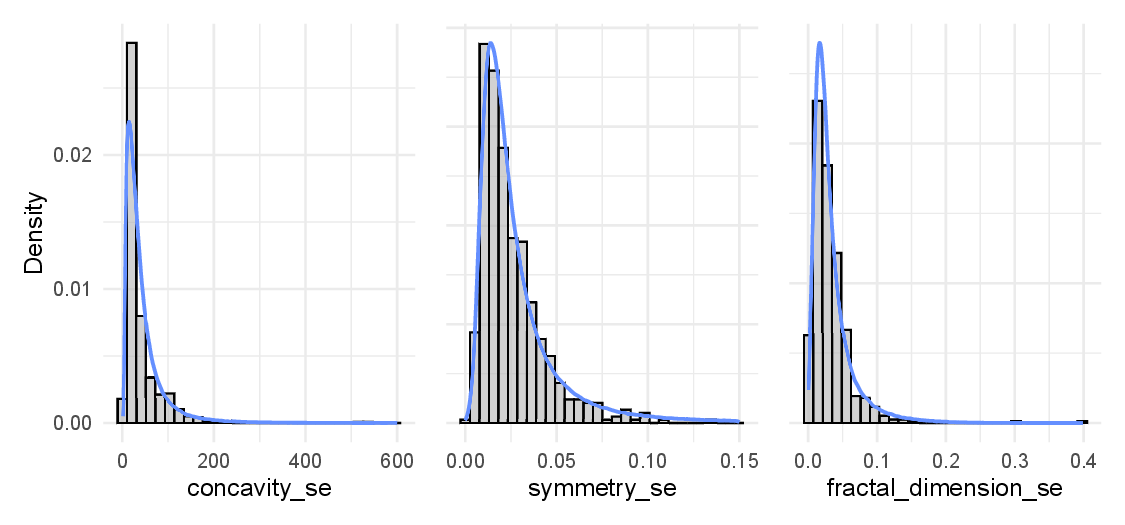}
    \caption{Marginal distributions of three tumor shape features from the observed dataset (histograms) overlaid with estimated densities (blue lines) based on an NCST model fitted using MLEs.}
    \label{fig:app_marginal}
\end{figure}

Now, we examine how well the NCST model fits the data using its maximum likelihood estimates (MLEs). Figure \ref{fig:app_marginal} presents the marginal distributions of the three tumor shape features from the observed data, overlaid with the corresponding marginal densities implied by the fitted NCST model. The plots show that the NCST model successfully captures the skewness and heavy-tailed behavior observed in the empirical distributions. The estimated density curves (blue lines) align closely with the histograms, especially for \texttt{symmetry\_se} and \texttt{fractal\_dimension\_se}, where the model provides a very strong visual fit across the majority of the range, effectively capturing the shape and concentration of the data. For \texttt{concavity\_se}, which exhibits more extreme right-skewness and heavier tails, the NCST model still captures the main body of the distribution effectively. However, there is some deviation in the extreme upper tail, where the model appears to slightly under-estimate the density of the most extreme observations. 

Overall, the agreement between the empirical histograms and model-based density estimates provides strong evidence that the NCST model, when fitted via MLE, offers a flexible and appropriate framework for capturing the marginal behavior of these biologically relevant tumor shape features.  

\begin{figure}[ht]
    \centering
    \includegraphics[width=\linewidth]{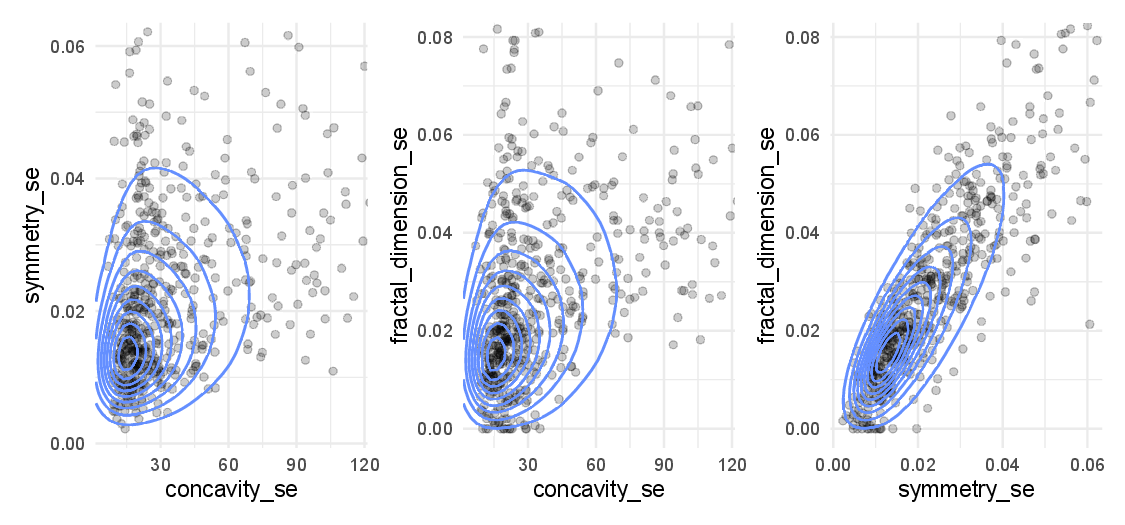}
    \caption{Zoomed contour plots of the NCST model fit. Blue contours represent the model-implied density based on MLEs, overlaid on observed tumor shape data. Axes are truncated at the 95th percentile to highlight regions with high data density.}
    \label{fig:app_contours}
\end{figure}

To further assess the fit of the NCST model, we examine its implied joint density structure using two-dimensional contour plots. Figure \ref{fig:app_contours} displays the observed data (black points) alongside contour curves based on kernel density estimates from a large NCST sample generated using the fitted MLEs. Each panel represents a different bivariate projection among the three tumor shape features. 

To improve interpretability and highlight the model's performance in regions of high data concentration, the axes in each panel are truncated at the 95th percentile of the observed values. This zoomed-in perspective focuses attention on the bulk of the data, where model fit is most consequential for practical applications. The contour plots reveal that the NCST model effectively captures the primary structure of the joint distributions. For example, the \texttt{symmetry\_se} vs. \texttt{fracal\_dimension\_se} panel shows tight alignment between the contour levels and the observed data cloud, suggesting a strong local fit. Similarly, the \texttt{\detokenize{fractal_dimension_se}}  vs. \texttt{\detokenize{symmetry_se}} projection reflects the positive skew and association present in the observed sample. 

The \texttt{concavity\_se} vs. \texttt{symmetry\_se} panel shows slightly looser alignment in the upper tail of \texttt{concavity\_se}, where the extreme observations lie outside the main contour region. This divergence is consistent with the heavy-tailed nature of that variable, but the model still captures the central data structure well. 

Overall, the contour plots provided additional support for the adequacy of the NCST model in capturing both marginal and joint patterns in the tumor shape data, even in the presence of skewness and heavy-tailed features. 

\section{Discussion and Conclusion}\label{sec:conclusion}
In this study, we introduced a multivariate non-central skew $t$ (NCST) distribution, constructed by scaling multivariate skew-normal random vectors with chi-squared distributed variables. This formulation yields a flexible family of distributions capable of capturing asymmetry, heavy tails, and non-centrality, which are features often observed in applied data but inadequately captured by standard multivariate distributions. 

We explore the theoretical properties of the NCST distribution, including moments, affine transformations, and the behavior of its quadratic forms. In particular, we established sufficient conditions under which quadratic forms of NCST variables follow a non-central skew $F$ distribution. The simulation studies confirmed this theoretical result by showing strong alignment between the empirical distribution of the quadratic forms and their theoretical counterparts under these conditions.

While the NCST model is more flexible, it encounters computational difficulties due to the lack of a closed-form likelihood. We addressed this through Monte Carlo likelihood approximation, enabling practical inference. However, this simulation-based approach exhibits sensitivity to the Monte Carlo sample size $M$, initial parameter values, and the choice of optimization method, sometimes leading to convergence at suboptimal local maxima. In our implementation, we observed improved stability when initializing the algorithm using estimates from Azzalini's skew $t$ distribution. Nonetheless, this sensitivity remains an inherent challenge of simulation-based inference, and it suggests that robust initialization and diagnostics are critical in practice. 

To evaluate the use of the NCST distribution in an applied setting, we analyzed a real-world dataset, focusing on tumor shape features extracted from diagnostic breast cancer data. Model comparisons using the SIC showed that the NCST model provided the best fit among several candidate distributions. Marginal and bivariate visualizations further demonstrated that the NCST model reasonably captures the skewed and heavy-tailed behavior in the data. This supports its suitability for modeling asymmetric biological features with complex dependence structures.

Looking forward, several research directions can enhance the utility of the NCST framework: 

\begin{itemize}
    \item Bayesian estimation methods may improve inference stability in small-sample or ill-\\ conditioned settings. 
    \item Quasi-Monte Carlo techniques or surrogate likelihood approximations could mitigate computational costs while maintaining accuracy.
    \item Further theoretical development of marginal and conditional distributions would support broader applications such as prediction and imputation. 
    \item Copula-based extensions could decouple marginal behavior from dependence structures, enabling richer modeling of asymmetric and heavy-tailed relationships. 
    \item Finally, integrating the NCST distribution into time series or longitudinal modeling frameworks, such as vector autoregressive or hierarchical models, offers a pathway to more realistic modeling of complex, temporally structured data. 
\end{itemize}

Together, these extensions will expand the applicability of the NCST model across a wide range of modern statistical problems where skewness and heavy tails are non-negligible.

%\textcolor{red}{We should cite some references for future research directions}

\bibliographystyle{apalike}  % Or another .bst style like apa, abbrvnat, etc.
\bibliography{MVST_refs}  % No .bib file extension needed

\begin{thebibliography}{}

\bibitem[Ahmad et~al., 2022]{ahmad2022new}
Ahmad, Z., Mahmoudi, E., and Dey, S. (2022).
\newblock A new family of heavy tailed distributions with an application to the
  heavy tailed insurance loss data.
\newblock {\em Communications in Statistics-Simulation and Computation},
  51(8):4372--4395.

\bibitem[Arellano-Valle, 2010]{ArellanoValle2010}
Arellano-Valle, R.~B. (2010).
\newblock On the information matrix of the multivariate skew-t model.
\newblock {\em Metron - International Journal of Statistics}, 0(3):371--386.

\bibitem[Arellano-Valle and Azzalini, 2013]{ARELLANOVALLE2013}
Arellano-Valle, R.~B. and Azzalini, A. (2013).
\newblock The centred parameterization and related quantities of the skew-t
  distribution.
\newblock {\em Journal of Multivariate Analysis}, 113:73--90.
\newblock Special Issue on Multivariate Distribution Theory in Memory of Samuel
  Kotz.

\bibitem[Azzalini, 1985]{Azzalini1985}
Azzalini, A. (1985).
\newblock A class of distributions which includes the normal ones.
\newblock {\em Scandinavian journal of statistics}, pages 171--178.

\bibitem[Azzalini, 2013]{Azzalini_2013}
Azzalini, A. (2013).
\newblock {\em The Skew-Normal and Related Families}.
\newblock Institute of Mathematical Statistics Monographs. Cambridge University
  Press.

\bibitem[Azzalini and Capitanio, 1999]{Azzalini1999}
Azzalini, A. and Capitanio, A. (1999).
\newblock Statistical applications of the multivariate skew normal
  distribution.
\newblock {\em Journal of the Royal Statistical Society. Series B (Statistical
  Methodology)}, 61(3):579--602.

\bibitem[Azzalini and Valle, 1996]{AzzaliniDallaValle1996}
Azzalini, A. and Valle, A.~D. (1996).
\newblock The multivariate skew-normal distribution.
\newblock {\em Biometrika}, 83(4):715--726.

\bibitem[Galarza et~al., 2022]{GALARZA2022104944}
Galarza, C.~E., Matos, L.~A., Castro, L.~M., and Lachos, V.~H. (2022).
\newblock Moments of the doubly truncated selection elliptical distributions
  with emphasis on the unified multivariate skew-t distribution.
\newblock {\em Journal of Multivariate Analysis}, 189:104944.

\bibitem[G{\'o}mez et~al., 2019]{gomez2019gumbel}
G{\'o}mez, Y.~M., Bolfarine, H., and G{\'o}mez, H.~W. (2019).
\newblock Gumbel distribution with heavy tails and applications to
  environmental data.
\newblock {\em Mathematics and Computers in Simulation}, 157:115--129.

\bibitem[Guo, 2017]{guo2017heavy}
Guo, Z.-Y. (2017).
\newblock Heavy-tailed distributions and risk management of equity market tail
  events.
\newblock {\em Journal of Risk \& Control}, 4(1):31--41.

\bibitem[Gupta, 2003]{Gupta2003}
Gupta, A.~K. (2003).
\newblock Multivariate skew $t$-distribution.
\newblock {\em Statistics: A Journal of Theoretical and Applied Statistics},
  37(4):359--363.

\bibitem[Hasan et~al., 2016]{HasanNingGupta2016}
Hasan, A., Ning, W., and Gupta, A. (2016).
\newblock A new approach for the skew $t$ distribution with applications to
  environmental data.
\newblock {\em Advances and Applications in Statistics}, 49(2):117--136.

\bibitem[Hasan, 2013]{Hasan2013}
Hasan, A.~M. (2013).
\newblock {\em A study of non-central skew $t$ distributions and their
  applications in data analysis and change point detection}.
\newblock PhD thesis, Bowling Green State University.

\bibitem[Ibragimov et~al., 2015]{ibragimov2015heavy}
Ibragimov, M., Ibragimov, R., and Walden, J. (2015).
\newblock {\em Heavy-tailed distributions and robustness in economics and
  finance}, volume 214.
\newblock Springer.

\bibitem[Johnson et~al., 1995]{johnson1995continuous}
Johnson, N.~L., Kotz, S., and Balakrishnan, N. (1995).
\newblock {\em Continuous Univariate Distributions}, volume~2 of {\em Wiley
  Series in Probability and Mathematical Statistics}.
\newblock Wiley-Interscience, New York.

\bibitem[Karlsson et~al., 2023]{karlsson2023vector}
Karlsson, S., Mazur, S., and Nguyen, H. (2023).
\newblock Vector autoregression models with skewness and heavy tails.
\newblock {\em Journal of Economic Dynamics and Control}, 146:104580.

\bibitem[Merz et~al., 2022]{merz2022understanding}
Merz, B., Basso, S., Fischer, S., Lun, D., Bl{\"o}schl, G., Merz, R., Guse, B.,
  Viglione, A., Vorogushyn, S., Macdonald, E., et~al. (2022).
\newblock Understanding heavy tails of flood peak distributions.
\newblock {\em Water Resources Research}, 58(6):e2021WR030506.

\bibitem[Nakajima, 2012]{Nakajima2012}
Nakajima, J. (2012).
\newblock Bayesian analysis of generalized autoregressive conditional
  heteroskedasticity and stochastic volatility: Modeling leverage, jumps and
  heavy-tails for financial time series.
\newblock {\em The Japanese Economic Review}, 63(1):81--103.

\bibitem[Tagle et~al., 2020]{TAGLE2020}
Tagle, F., Castruccio, S., and Genton, M.~G. (2020).
\newblock A hierarchical bi-resolution spatial skew-t model.
\newblock {\em Spatial Statistics}, 35:100398.

\bibitem[Wang et~al., 2009]{WANG2009533}
Wang, T., Li, B., and Gupta, A.~K. (2009).
\newblock Distribution of quadratic forms under skew normal settings.
\newblock {\em Journal of Multivariate Analysis}, 100(3):533--545.

\bibitem[Wolberg et~al., 1993]{breast_cancer_wisconsin_(diagnostic)_17}
Wolberg, W., Mangasarian, O., Street, N., and Street, W. (1993).
\newblock {Breast Cancer Wisconsin (Diagnostic)}.
\newblock UCI Machine Learning Repository.
\newblock {DOI}: https://doi.org/10.24432/C5DW2B.

\end{thebibliography}

%\printbibliography
%\newpage
%\section{Appendix: Related Definition and Results}\label{sec:Appendix}
%\input{09_Appendix}
\end{document}